\documentclass[showpacs, secnumarabic, amssymb, nobibnotes, aip, jap, reprint]{revtex4-1}
\usepackage{graphicx}%
\usepackage{dcolumn}%
\usepackage{bm}%
\usepackage{lineno}

\begin{document}

\title{Negative magnetoresistance and anomalous Hall effect \\ in GeMnTe-SnMnTe spin-glass-like system}

\author{L.~Kilanski}
\email[Electronic mail: ]{kilan@ifpan.edu.pl}
\author{R.~Szymczak}
\author{W.~Dobrowolski}
\author{A.~Podg\'{o}rni}
\author{A.~Avdonin}
\affiliation{Institute of Physics, Polish Academy of Sciences, al. Lotnikow 32/46, 02-668 Warsaw, Poland}

\author{V.~E.~Slynko}
\author{E.~I.~Slynko}
\affiliation{Institute of Materials Science Problems, Ukrainian Academy of Sciences, 5 Wilde Street, 274001 Chernovtsy, Ukraine}

\date{\today}

\begin{abstract}

Magnetotransport properties of spin-glass-like Ge$_{1\textrm{-}x\textrm{-}y}$Sn$_{x}$Mn$_{y}$Te mixed crystals with chemical composition changing in the range of 0.083$\,$$\leq$$\,$$x$$\,$$\leq$$\,$0.142 and 0.012$\,$$\leq$$\,$$y$$\,$$\leq$$\,$0.119 are presented. The observed negative magnetoresistance we attribute to two mechanisms i.e. weak localization occurring at low fields and spin disorder scattering giving contribution mainly at higher magnetic fields. A pronounced hysteretic anomalous Hall effect (AHE) was observed. The estimated AHE coefficient shows a small temperature dependence and is dependent on Mn-content, with changes in the range of 10$^{-7}$$\,$$<$$\,$$R_{S}$$\,$$<$$\,$10$^{-6}$$\;$m$^{3}$/C. The scaling law analysis has proven that the AHE in this system is due to the extrinsic mechanisms, mainly due to the skew scattering accompanied with the side jump processes.

\end{abstract}

\keywords{spintronics; semimagnetic-semiconductors; ferromagnetic-materials; electronic-transport}

\pacs{72.80.Ga, 75.40.Cx, 75.40.Mg, 75.50.Pp}



\maketitle

\linenumbers

\section{Introduction}

Semiconductor spintronics is being intensively studied for the last two decades. Magnetic order due to carrier induced magnetic interactions was observed in many conventional III-V, II-VI and IV-VI compound semiconductors such as transition metal doped PbSnTe, GaAs and other diluted magnetic semiconductors.\cite{Kossut1993a, Matsukura2002a, Dobrowolski2003a} The presence of carrier induced magnetic interactions with room temperature magnetic ordering is needed for making use of diluted magnetic semiconductors in semiconductor spintronics. The Curie temperature of the most intensively studied and technologically mastered semimagnetic semiconductor Ga$_{1\textrm{-}x}$Mn$_{x}$As does not exceed 185$\;$K (Ref.$\;$\onlinecite{Wang2008a}), which excludes the practical application of this material. It is therefore necessary to look for alternative compounds that can operate at room temperature. \\ \indent Semimagnetic semiconductors based on IV-VI group of periodic table, in particular Ge$_{1\textrm{-}x}$TM$_{x}$Te alloys (TM - transition metal) are perspective and intensively studied materials\cite{Hassan11a, Knoff09a, Fukuma08a} due to appearance of carrier mediated ferromagnetism with high Curie temperatures reaching 200$\;$K in Ge$_{1\textrm{-}x}$Mn$_{x}$Te with $x$$\,$$=$$\,$0.46 (see Ref.$\;$\onlinecite{Lechner2010a}). GeTe is a narrow gap semiconductor with $E_{g}$$\,$$=$$\,$0.23$\;$eV (Ref.$\;$\onlinecite{Edwards2006a}) crystallizing in rhombohedrally distorted NaCl structure. Ge$_{1\textrm{-}x}$TM$_{x}$Te alloys can be considered as multiferroics, since ferroelectric order is introduced via rhombohedral distortion. Negative magnetoresistance \cite{Brodowska2008a} and anomalous Hall effect \cite{Brodowska2006a} are usually significant and widely observed in these materials. It is therefore necessary to bring this subject into considerable attention. Moreover, alloying of GeTe with SnTe should cause the alloy to change its electrical and optical properties, which is important in view of possible control of magnetic properties of IV-VI based semimagnetic semiconductors.  \\ \indent The present paper extends our previous investigation of structural and magnetic properties of GeMnTe-SnMnTe system\cite{Kilanski2010a, Kilanski2009a, Kilanski2008a} by an extensive study of magnetotransport properties. In this paper, we have made an analysis of the negative magnetoresistance occurring in the GeMnTe-SnMnTe system below the temperature of the transition to the spin glass state, $T_{SG}$. This effect can be well described by the existing theory of the spin-disorder scattering magnetoresistance and can be correlated with the magnetization of the studied material. Additionally, we have found a strong anomalous Hall effect (AHE), showing hysteresis in our samples. The estimated values of AHE coefficient, $R_{S}$, show a weak temperature dependence at $T$$\,$$\ll$$\,$$T_{SG}$, at the same time they strongly depend on the chemical composition of the samples.

\section{Sample characterization}

The samples being the subject of the current research are bulk crystals grown using a modified Bridgman method. The modifications of the growth procedure are similar to those applied by Aust and Chalmers for the growth of alumina crystals \cite{Aust1958a} and consist of the installation inside the growth furnace of additional heating elements creating a radial temperature gradient. It allows the modification of the slope of the crystallization plane by about 15 deg. The used modifications were proven as an effective tool for decreasing the number of the crystal blocks in the as grown ingots from a few down to one or two. \\ \indent The as grown ingots were cut into thin slices (typically around 1$\;$mm thick) perpendicular to the growth direction with the use of a precision wire saw. The chemical composition of each slice was determined with the use of energy dispersive x-ray spectroscopy (EDXRF).
\begin{table}
\caption{\label{TabPodstCharGeSnMnTe}Results of a basic characterization of Ge$_{1\textrm{-}x\textrm{-}y}$Sn$_{x}$Mn$_{y}$Te samples including the chemical composition $x$ and $y$, the Hall carrier concentration $n$ (measured at $T$$\,$$=$$\,$300$\;$K), and the spin-glass transition temperature $T_{SG}$.}
\begin{ruledtabular}
\begin{tabular}{cccc}

    $x$     & $y$   &  n [$10^{21}$$\;$cm$^{-3}$] &  $T_{SG}$ [K]   \\ \hline

   0.105$\,$$\pm$$\,$0.01   &  0.012$\,$$\pm$$\,$0.001   &  1.3$\pm$0.1  &  9.78$\pm$0.06   \\
   0.112$\,$$\pm$$\,$0.01   &  0.031$\,$$\pm$$\,$0.01    &  1.4$\pm$0.1  &  42.12$\pm$0.12   \\
   0.119$\,$$\pm$$\,$0.01   &  0.031$\,$$\pm$$\,$0.01    &  1.5$\pm$0.1  &  19.97$\pm$0.19   \\
   0.142$\,$$\pm$$\,$0.01   &  0.034$\,$$\pm$$\,$0.01    &  1.8$\pm$0.1  &  21.15$\pm$0.04   \\
   0.090$\,$$\pm$$\,$0.009  &  0.039$\,$$\pm$$\,$0.004   &  1.3$\pm$0.1  &  41.04$\pm$0.13   \\
   0.094$\,$$\pm$$\,$0.01   &  0.079$\,$$\pm$$\,$0.008   &  1.1$\pm$0.1  &  45.20$\pm$0.23  \\
   0.091$\,$$\pm$$\,$0.009  &  0.094$\,$$\pm$$\,$0.009   &  3.3$\pm$0.2  &  34.46$\pm$1.01   \\
   0.091$\,$$\pm$$\,$0.009  &  0.115$\,$$\pm$$\,$0.01    &  4.1$\pm$0.2  &  30.36$\pm$0.97   \\

\end{tabular}
\end{ruledtabular}
\end{table}
The maximum relative errors of the EDXRF technique does not exceed 10\% of the calculated value of $x$ or $y$. The EDXRF data shows a continuous change of the chemical composition of the slices along the growth direction. Among all the slices only a few have been selected, which are featured by: (i) having the lowest relative inhomogeneity within an individual slice and (ii) having Sn and Mn content covering the widest possible range of chemical compositions. From all our samples we selected a few (see Table$\;$\ref{TabPodstCharGeSnMnTe}) that had chemical composition changing in the range of 0.09$\,$$\leq$$\,$$x$$\,$$\leq$$\,$0.142 and 0.012$\,$$\leq$$\,$$y$$\,$$\leq$$\,$0.115. \\ \indent The powder x-ray diffraction (XRD) measurements were performed at room temperature. The results show that all our samples are single phased and are crystallized in rhombohedrally distorted NaCl structure, similarly to the binary nonmagnetic analog of our material, namely the GeTe compound. The XRD data analysis was done with the use of Rietveld method and it shows that the samples have lattice parameter $a$$\,$$\approx$$\,$5.98$\;$$\textrm{\AA}$ and the angle of rhombohedral distortion $\alpha$$\,$$\approx$$\,$88.3$^{\circ}$. These are similar values to those well established for GeTe system.\cite{Galazka1999a} It should be noted that the lattice parameter is a decreasing function of the Sn or Mn amount in the sample. However, since we have two different substitutional ions in the alloy, it is difficult to perform a detailed analysis of the results. \\ \indent The magnetic properties of our Ge$_{1\textrm{-}x\textrm{-}y}$Sn$_{x}$Mn$_{y}$Te samples were studied extensively and the details can be found in  Refs.$\;$\onlinecite{Kilanski2010a, Kilanski2009a, Kilanski2008a}. The main conclusions drawn from our previous investigation are the following:
\begin{itemize}
  \item All of the studied samples show magnetic transition at temperatures below 50$\;$K. The ac-susceptibility      studies revealed that the spin-glass-like state was observed with a transition temperature, $T_{SG}$, generally increasing as a function of the Mn content 0.012$\,$$\leq$$\,$$x$$\,$$\leq$$\,$0.115 and the carrier concentration 1$\times$10$^{21}$$\,$$<$$\,$$n$$\,$$<$$\,$4$\times$10$^{21}$$\;$cm$^{-3}$ in the range of 10$\,$$\leq$$\,$$T$$\,$$\leq$$\,$50$\;$K. The long-range RKKY interaction was found to be the leading physical mechanism responsible for the observed magnetic order.
  \item A well defined hysteresis loop was observed in all our spin-glass-like samples, indicating that the system was not an ideal spin-glass, but consisted of the ferromagnetic regions at which spin-glass freezing occurs for $T$$\,$$<$$\,$$T_{SG})$.
  \item The nonsaturating $M$($B$) magnetization curves were observed for $T$$\,$$<$$\,$$T_{SG}$ indicating the presence of strong magnetic frustration in our samples.
\end{itemize}

\section{Magnetotransport studies}

The magnetotransport studies of the Ge$_{1\textrm{-}x\textrm{-}y}$Sn$_{x}$Mn$_{y}$Te samples were performed in the standard dc-current six-contact Hall geometry. We have used the superconducting magnet with maximum magnetic field equal to $B$$\,$$=$$\,$13$\;$T and a sweep speed of about 0.5$\;$T/min, equipped with the cryostat allowing the control of the temperature of the sample in the range of 1.4$\,$$\leq$$\,$$T$$\,$$\leq$$\,$300$\;$K. The samples, cut to size of about 1$\times$1$\times$10$\;$mm, were etched and cleaned before making electrical connections. The contacts were made with the use of gold wire and indium solder. The ohmic behavior of each contact pair was checked prior to proper measurements. The magnetoresistance and the Hall effect were measured simultaneously at selected temperatures, covering temperatures both below and above magnetic phase transition in the samples.

\subsection{Negative Magnetoresistance}

\noindent The isothermal magnetoresistance measurements were performed for all our Ge$_{1\textrm{-}x\textrm{-}y}$Sn$_{x}$Mn$_{y}$Te samples. The $\rho_{xx}$($B$) curves were obtained by averaging the results for positive and negative current. In order to allow simple data presentation the $\rho_{xx}$($B$) curves at different temperatures were normalized to the zero-field resistivity value $\rho_{0}$ by using the following relation $\Delta \rho_{xx}/ \rho_{xx}(0)$$\,$$=$$\,$$(\rho_{xx}(B)$$\,$$-$$\,$$\rho_{xx}(B=0))$/$\rho_{xx}(B=0)$.
\begin{figure}[!h]
  \begin{center}
    \includegraphics[width = 0.5\textwidth, bb = 90 170 730 510]
    {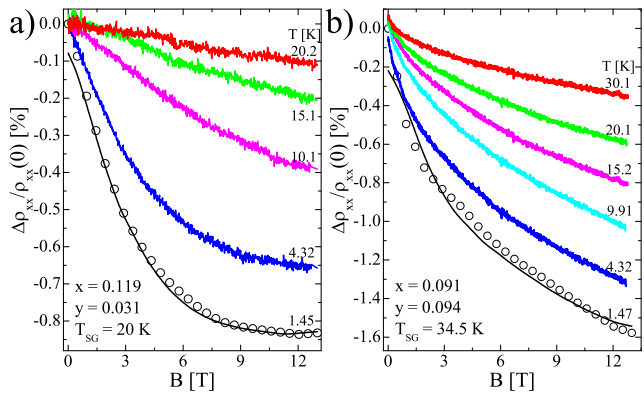}\\
  \end{center}
  \caption{\small Magnetoresistance curves obtained at different temperatures (for $T$$\,$$=$$\,$1.4$\;$K experimental data is marked by symbols and theoretical curve is depicted by line) for exemplary Ge$_{1\textrm{-}x\textrm{-}y}$Sn$_{x}$Mn$_{y}$Te samples with two different chemical compositions.}
  \label{FigMResCurves}
\end{figure}
The experimental data shows that for all our samples below the spin-glass transition temperature $T_{SG}$ the negative magnetoresistance is observed (exemplary results shown in Fig.$\;$\ref{FigMResCurves}). The magnetoresistance curves at $T$$\,$$<$$\,$$T_{SG}$ have negative value without saturation up to the maximum magnetic fields (equal to $B$$\,$$=$$\,$13$\;$T) used in our experiments. On the other hand at $T$$\,$$>$$\,$$T_{SG}$ only positive, classical orbital magnetoresistance with small amplitudes (maximum 0.1\%) was observed in all our samples. The negative magnetoresistance observed at $T$$\,$$\leq$$\,$$T_{SG}$ is isotropic. Our results indicate that the observed negative magnetoresistance is due to the influence of the magnetic impurities (present in this system) on the carrier transport in the presence of magnetic field. This conclusion may be justified by the data gathered in Fig.$\;$\ref{FigMResCurves}, where both the magnitude of the observed magnetoresistance and the spin-glass transition temperatures, $T_{SG}$ are found to be strongly correlated with the amount of Mn, $y$.
\begin{figure}[!h]
  \begin{center}
    \includegraphics[width = 0.42\textwidth, bb = 0 40 655 625]
    {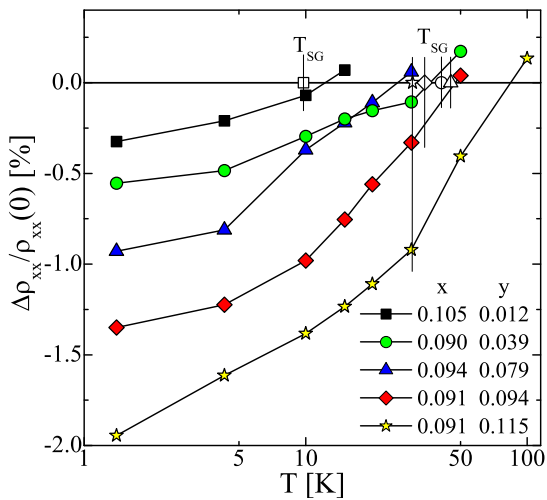}\\
  \end{center}
  \caption{\small The amplitude of the magnetoresistance observed in the studied Ge$_{1\textrm{-}x\textrm{-}y}$Sn$_{x}$Mn$_{y}$Te samples with different chemical composition. The open symbols represents the spin-glass transition temperatures, as obtained from the magnetometric measurements in Ref.$\;$\onlinecite{Kilanski2010a}.}
  \label{FigMResAmpl}
\end{figure}
Inspection of Fig.$\;$\ref{FigMResAmpl} shows that the magnitude of the magnetoresistance obtained at $T$$\,$$\approx$$\,$1.5$\;$K is a nearly linear function of Mn content, $y$. Moreover, the negative magnetoresistance is diminishing at $T$$\,$$>$$\,$$T_{SG}$ for most of our samples, except the crystals with the highest Mn content $y$$\,$$\geq$$\,$0.094. We connect this discrepancy with a significant magnetic frustration of the material (which is the strongest in samples with high manganese content) makes the transition process to extend over $T_{SG}$. However, a detailed data analysis needs to be performed in order to clarify the physical mechanism responsible for the negative magnetoresistance in our material. \\ \indent A number of different physical phenomena might be responsible for the negative magnetoresistance of a conductor doped or alloyed with magnetic impurities. Weak localization phenomenon \cite{Anderson1958a} is commonly attributed to be the mechanism leading to the negative magnetoresistance at low temperatures. However, this effect should diminish at relatively high magnetic fields used in our experiments (in our experiments at $B$$\,$$\approx$$\,$13$\;$T the negative magnetoresistance does not show saturation), where the constructive interference of the wave functions of the free-carriers and Mn-impurity d-electrons cannot further diminish. The appearance of negative magnetoresistance is usually connected in spin-glasses with the strong sp-d exchange coupling.\cite{Majumdar1983a} Since our system shows features characteristic for both spin-glass \cite{Kilanski2009a} and ferromagnetic \cite{Kilanski2010a} materials we should consider its magnetic order to be similar to mictomagnetic order, where the spin-glass frustration is accompanied by microscopic regions where the domain structure is formed for $T$$\,$$\leq$$\,$$T_{SG}$. The magnetoresistance of disordered spin-glass should follow the general scaling relation $\rho_{xx}$$\,$$\propto$$\,$-$\alpha$$M^{2}$, where $\alpha$ is a proportionality constant.\cite{Majumdar1983a} However, in our case the magnetoresistance does not scale with the magnetization according to the above relation. Thus, we can conclude that the magnetoresistance in our system probably has a different origin than it was proposed for canonical spin-glasses. \\ \indent The amplitude of the negative magnetoresistance observed in our samples is similar to that reported for Ge$_{1\textrm{-}x}$Mn$_{x}$Te layers \cite{Fukuma2002a} and is most probably due to the reduction of spin-disorder in the presence of an applied external magnetic field. This is well justified by the fact, that the amplitude of magnetoresistance in our samples is proportional to the amount of Mn, $y$. According to de Gennes and Fisher\cite{Gennes1958a, Fisher1968a} the reduction of the carrier scattering on paramagnetic moments due to the application of the static magnetic field can be expressed using the following relation
\begin{equation}\label{EqMRSpDiScat}
    \rho_{sd} = 2 \pi^{2} \frac{k_{F}}{ne^{2}} \frac{m^{2} \Gamma_{S}^{2}}{h^{3}} n_{S} \Big{[} S(S+1) - \langle S \rangle^{2}_{B,T} \Big{]},
\end{equation}
where $\rho_{sd}$ is the contribution to the resistivity resulting from the spin disorder scattering mechanism, $e$ is the elementary charge, $k_{F}$ is the Fermi wave vector, $m$ is the electron mass, $h$ is the Planck constant, $n_{S}$ is the density of 3d electrons of paramagnetic ions, $\Gamma_{S}$$\,$$=$70$\;$eVA$^{3}$ (value taken for Ga$_{1\textrm{-}x}$Mn$_{x}$As from Ref.$\;$\onlinecite{Yoon2004a}) is an effective factor related to the conducting carrier - magnetic ion exchange integral and $S$$\,$$=$$\,$5/2 is the spin quantum number of the Mn ion. For the system with spin-only ground state Eq.$\;$\ref{EqMRSpDiScat} can be rewritten in the following form
\begin{widetext}
\begin{equation}\label{EqMRSpDiScat2}
    \rho_{sd} = 2 \pi^{2} \frac{k_{F}}{ne^{2}} \frac{m^{2} \Gamma_{S}^{2}}{h^{3}} n_{S} \Bigg{\langle} \frac{1}{2} + \Bigg{[} \exp{\Big{(}\frac{-g_{s} \mu_{B} \mu_{0} B}{2k_{B}T}\Big{)}} + \exp{\Big{(}\frac{g_{s} \mu_{B} \mu_{0} B}{2k_{B}T}\Big{)}} \Bigg{]}^{-2}  \Bigg{]},
\end{equation}
\end{widetext}
where $g_{S}$ is an effective factor (related to the average effective magnetic moment per Mn ion), $\mu_{B}$ is Bohr magneton and $B$ is the amplitude of the external magnetic field. The values of parameters in Eq.$\;$\ref{EqMRSpDiScat2} were estimated from other experimental results. The $g_{S}$ parameter was the only fitting parameter. We attempted to fit the experimental results, assuming that there exists a positive, square contribution to the magnetoresistance in our system, associated with orbital motion of conducting carriers in the magnetic field. The resulting theoretical curves describe the experimental results only for magnetic fields $B$$\,$$>$$\,$1$\;$T. This signifies that at low magnetic fields, another contribution to the negative magnetoresistance is present. It is very likely that the weak localization of carriers on the defect states of the crystal lattice is the source of this additional contribution to the magnetoresistance. For the above reasons, we repeated the fitting to the Eq.$\;$\ref{EqMRSpDiScat2}, limiting it to 1$\,$$<$$\,$$B$$\,$$<$$\,$13$\;$T. Our analysis was done for the lowest measurement temperatures $T$$\,$$\approx$$\,$1.4$\;$K, where variances of the fitting parameters had the smallest values (due to largest amplitudes of the magnetoresistance). The theoretical curves obtained in this way reproduce the experimental results much better. As a result of the data analysis we have estimated the $g_{S}$ values, which were similar for all our samples and temperatures and equal to $g_{S}$$\,$$\approx$$\,$4.0$\pm$0.5 at $T$$\,$$\approx$$\,$1.4$\;$K (see lines in Fig.$\;$\ref{FigMResCurves}). The obtained values of $g_{S}$ provide value of the magnetic moment $m$$\,$$\approx$$\,$2$\;$$\mu_{B}$/Mn ion. The obtained magnetic moment values are significantly lower than the corresponding value of $m$$\,$$=$$\,$5$\;$$\mu_{B}$/Mn ion for Mn$^{2+}$ with $S$$\,$$=$$\,$5/2. These results are consistent with the previous estimates carried out on the basis of the results of magnetometric measurements (see Ref.$\;$\onlinecite{Kilanski2010a}), which yielded in a much smaller magnetic moment of Mn ion in Ge$_{1\textrm{-}x\textrm{-}y}$Sn$_{x}$Mn$_{y}$Te samples. This confirms our earlier findings that the distribution of Mn ions in the GeTe crystal lattice is far from being perfect. The presence of antiferromagnetic substitutional-interstitial Mn pairs is highly probable in our system which causes a large fraction of Mn ions to be magnetically inactive. Such effect is well known in semimagnetic semiconductors, in particular in Ga$_{1\textrm{-}x}$Mn$_{x}$As layers,\cite{Kacman2005a} where the antiferromagnetic Mn pairs lower the effective magnetic moment of entire system of Mn ions.

\subsection{Anomalous Hall Effect}

\begin{figure}[!h]
  \begin{center}
    \includegraphics[width = 0.5\textwidth, bb = 70 110 750 500]
    {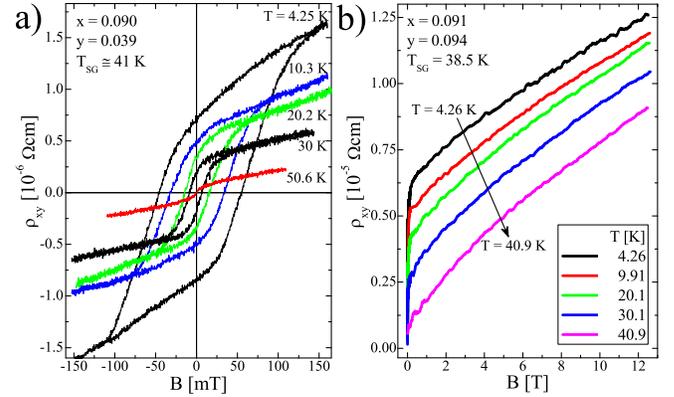}\\
  \end{center}
  \caption{\small Results of the Hall effect measurements for selected Ge$_{1\textrm{-}x\textrm{-}y}$Sn$_{x}$Mn$_{y}$Te samples with different chemical composition (see legends) showing strong AHE including (a) hysteretic behavior of the isothermal $\rho_{xy}$($B$) curves and (b) high field Hall effect showing strong AHE.}
  \label{FigAHECurves}
\end{figure}

\begin{figure}[!h]
  \begin{center}
    \includegraphics[width = 0.42\textwidth, bb = 0 40 640 580]
    {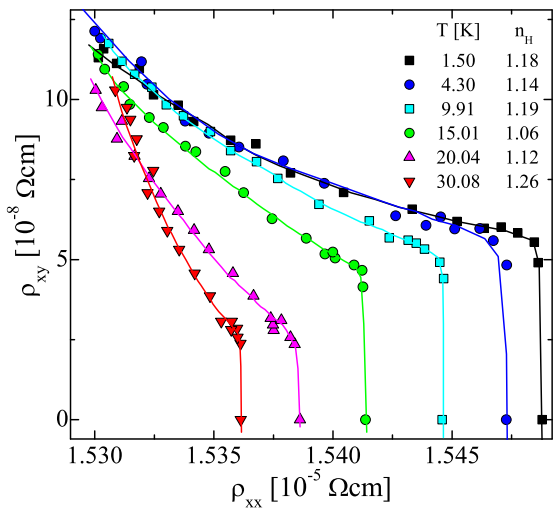}\\
  \end{center}
  \caption{\small The Hall resistivity component $\rho_{xy}$ as a function of the parallel resistivity component $\rho_{xx}$ obtained experimentally (points) at a few temperatures for selected Ge$_{0.815}$Sn$_{0.091}$Mn$_{0.094}$Te sample and fitted (lines) to the scaling relation given by Eq.$\;$\ref{EqAHEScal01}. Different points correspond to different values of magnetic field. }
  \label{FigAHEScal}
\end{figure}

\noindent The magnetic field dependencies of the resistivity component perpendicular to the current and magnetic field direction, namely $\rho_{xy}$($B$), was measured at several stabilized temperatures below, near and above the spin-glass transition temperatures $T_{SG}$. Our results indicate clearly that, for all Ge$_{1\textrm{-}x\textrm{-}y}$Sn$_{x}$Mn$_{y}$Te samples, below $T_{SG}$, the $\rho_{xy}$($B$) curves show strong anomalous Hall effect (AHE) and hysteresis. The exemplary results of the Hall effect measurements for selected Ge$_{1\textrm{-}x\textrm{-}y}$Sn$_{x}$Mn$_{y}$Te samples are presented in Fig.$\;$\ref{FigAHECurves}.
The comparison of the magnetometric (not shown here - for details see Ref.$\;$\onlinecite{Kilanski2010a}) and magnetotransport data shows that the coercive fields obtained from both types of measurements coincide with a good accuracy. This indicates that in our spin-glass-like system occurs the asymmetric carrier scattering and it can be directly linked to the magnetic properties of the alloy. The selected Hall effect curves presented in Fig.$\;$\ref{FigAHECurves}b show that the AHE makes a significant contribution to the total Hall effect in this system at $T$$\,$$<$$\,$$T_{SG}$. As can be seen, the Hall effect curves show no linearity even at the highest magnetic fields used during the measurements i.e. up to $B$$\,$$=$$\,$13$\;$T. This feature is related with the lack of the saturation of magnetization in our samples (data not shown here, for details - see Ref.$\;$\onlinecite{Kilanski2010a}). In order to quantify the strength of the AHE and to estimate the Hall carrier concentration and mobility for $T$$\,$$<$$\,$$T_{SG}$ an appropriate fitting procedure must be employed. \\ \indent The Hall effect in a conductor doped with magnetic ions, in its magnetically ordered temperature region, shows the usual Lorentz term $R_{H}B$ and a second contribution, namely AHE, caused by the asymmetric carrier scattering. The AHE is due to the spin-orbit coupling in the presence of spin-polarization (for details see Ref.$\;$\onlinecite{Sinova2004a} and references therein). The AHE term in some cases dominates the total Hall effect below the Curie temperature, thus making the precise estimation of the carrier concentration and mobility very difficult. The magnetic field dependence of the Hall resistivity tensor component $\rho_{xy}$  in the standard six contact Hall geometry can be expressed using the following relation
\begin{equation}\label{EqAHE}
    \rho_{xy}(B) = R_{H} B + \mu_{0} R_{S} M,
\end{equation}
where $R_{H}$ and $R_{S}$ are the normal and anomalous Hall coefficients, $\mu_{0}$ is the magnetic permeability constant, and $M$ is the magnetization of the sample. Both the ordinary and anomalous Hall coefficient can be extracted from the total Hall effect with the knowledge about the magnetic field dependence of the magnetization at given temperatures. The use of the $M$($B$) curve is crucial especially for a system in which the magnetization does not show saturation even at relatively high fields $B$$\,$$=$$\,$9$\;$T. In such a system the AHE term gives a contribution that is not constant as a function of the magnetic field and not only the ordinary term of the Hall effect affects the $\rho_{xy}$(B) dependence and causes it to be an increasing function of the applied magnetic field. Thus, an elaborated fitting procedure needs to be employed in order to quantify the Hall effect data i.e. to precisely calculate the $R_{H}$ and $\mu$ at low temperatures $T$$\,$$<$$\,$$T_{SG}$. \\ \indent In order to properly quantify the strength of the observed AHE and to calculate the Hall constant and carrier mobility a fitting of the data to the Eq.$\;$\ref{EqAHE} was performed. The least square fits of the experimental magnetic field dependencies of the off-diagonal resistivity tensor component $\rho_{xy}$($B$,$T$) and the isothermal magnetization curves $M$($B$) to the Eq.$\;$\ref{EqAHE} were performed. The fitting procedure for the $\rho_{xy}$($B$,$M$($B$))$|_{T=const}$ function was done with the use of Minuit functional minimalization package\cite{James1975a} in two steps. At first, both the ordinary and anomalous Hall constants were taken as the fitting parameters. The first series of least-square fits gave similar values of the ordinary Hall constant $R_{H}$. This is a reasonable result, since our samples show a metallic-like resistivity vs. temperature dependence. Thus, since no thermal activation of the conducting holes to the valence band occurred, one should not observe any temperature dependence of the Hall carrier concentration. After the first series of fits was done for the data acquired at several constant temperatures the average value of the $R_{H}$ was calculated. The Hall carrier concentrations obtained from the average value of $R_{H}$ (see Table$\;$\ref{TabAHEFitRes}) were around $n$$\,$$\approx$$\,$10$^{21}$$\;$cm$^{-3}$, which is a value typical of GeTe based semiconductors. The low temperature carrier mobility was found to have rather low values $\mu$$\,$$<$$\,$15$\;$cm$^{2}$/(Vs). \\ \indent During the second series of fits only the anomalous Hall constant $R_{S}$ was taken as a fitting parameter. The obtained values of $R_{S}$ presented as as a function of temperature for the studied Ge$_{1\textrm{-}x\textrm{-}y}$Sn$_{x}$Mn$_{y}$Te samples with different chemical composition (see legends) do not show any large temperature dependence. The average values of $R_{S}$ obtained for our samples are gathered in Table$\,$\ref{TabAHEFitRes}.
\begin{table}
\caption{\label{TabAHEFitRes} Results of the fitting of the experimental Hall effect data to Eqs$\;$\ref{EqAHE} and \ref{EqAHEScal01} including the low temperature (valid for $T$$\,$$<$$\,$$T_{SG}$) estimate of the Hall constant $R_{H}$, the Hall carrier mobility $\mu$, the anomalous Hall constant $R_{S}$, and the scaling coefficient $n_{H}$. The errors were calculated as mean square deviation.}
\begin{ruledtabular}
\begin{tabular}{cccccc}

    $x$     & $y$   &  $R_{H}$               &  $\mu$                 &  $R_{S}$                &  $n_{H}$  \\
            &       &  [$10^{-9}$ m$^{3}$/C] &  cm$^{2}$/(V$\cdot$s)  &  [$10^{-7}$ m$^{3}$/C]  &           \\ \hline

   0.105    &   0.012   &  7.2$\pm$0.6  &  7.0$\pm$0.2   &  7.2$\pm$0.5  &  1.2$\pm$0.1  \\
   0.112    &   0.031   &  6.5$\pm$0.4  &  5.0$\pm$0.5   &   19$\pm$2    &  1.2$\pm$0.1  \\
   0.119    &   0.031   &  6.0$\pm$0.3  &  4.2$\pm$0.5   &  9.5$\pm$0.5  &  1.2$\pm$0.1  \\
   0.142    &   0.034   &  6.7$\pm$0.4  &  4.0$\pm$0.3   &  5.3$\pm$0.4  &  1.3$\pm$0.1  \\
   0.090    &   0.039   &  8.1$\pm$0.6  &   25$\pm$2     &  5.0$\pm$0.3  &  1.2$\pm$0.1  \\
   0.094    &   0.079   &  6.0$\pm$0.4  &  5.3$\pm$0.4   &  4.2$\pm$0.3  &  1.1$\pm$0.1  \\
   0.091    &   0.094   &  3.2$\pm$0.3  &  3.0$\pm$0.2   &   11$\pm$1    &  1.2$\pm$0.1  \\
   0.091    &   0.115   &  8.3$\pm$0.5  &   14$\pm$1     &  9.7$\pm$0.8  &  1.1$\pm$0.1  \\

\end{tabular}
\end{ruledtabular}
\end{table}
The values of $R_{S}$ obtained in this work are higher than the ones reported for other IV-VI based diluted magnetic semiconductors such as Sn$_{1\textrm{-}x\textrm{-}y}$Mn$_{x}$Er$_{y}$Te and Ge$_{1\textrm{-}x\textrm{-}y}$Mn$_{x}$Eu$_{y}$Te.\cite{Brodowska2006a, Brodowska2008a} The $R_{S}$ values indicate that in the case of Ge$_{1\textrm{-}x\textrm{-}y}$Sn$_{x}$Mn$_{y}$Te crystals, in which the Mn content was smaller than $y$$\,$$=$$\,$0.05, there is a relationship between the chemical composition and the values of $R_{S}$. There is a drop in the $R_{S}$ with the increase of the amount of Sn ions in the alloy. The observed trends in $R_{S}$ with both $x$ and $y$ are similar to the trends in the coercive field $H_{C}$ with the amount of Sn and Mn (not shown here, for details see Ref.$\;$\onlinecite{Kilanski2010a}), and therefore a change of the domain structure of the material, which could have a significant impact on the asymmetric scattering of carriers, leading to the AHE. It should be noted, that no significant temperature dependence of the AHE coefficient $R_{S}$ was observed, in agreement with the results reported for other IV-VI semiconductors \cite{Brodowska2006a}. A strong decreasing $R_{S}$($T$) dependence was observed in only two of our samples i.e. the crystals with $x$$\,$$=$$\,$0.090, $y$$\,$$=$$\,$0.039 and $x$$\,$$=$$\,$0.091, $y$$\,$$=$$\,$0.115. The reason for this decrease is not understood. The Hall carrier concentration for these two samples is the lowest and the Hall carrier mobility is the highest among all our samples, which might have a major influence on the carrier scattering (since lower carrier concentration results from a smaller amount of cation vacancies in this sample) in the material and, consequently, on the AHE. In a second group of Ge$_{1\textrm{-}x\textrm{-}y}$Sn$_{x}$Mn$_{y}$Te crystals, i.e. those with a high Mn content, the evident increase in the $R_{S}$($y$) dependence with the increasing $y$ was observed. These changes are also correlated with the $H_{C}$($y$) relationship, which is a decreasing function of $y$. We can speculate that this could mean that both values are somewhat related. This conclusion may be supported by the fact that in both groups of crystals a general reduction of $R_{S}$ with an increase of the coercive field of the crystal was observed, and therefore the changes of the domain structure of the material are likely to be critical for explaining the AHE in this material. \\ \indent It is a fact well known in the literature, that there are two major mechanisms leading to the formation of AHE, namely skew scattering and side jump, which can be described theoretically and distinguished by appropriate linear \cite{Smit1955a} and square \cite{Berger1970a} dependencies between the resistivity components $\rho_{xy}$$\,$$\propto$$\,$$\rho_{xx}^{n_{H}}$, 1$\,$$\leq$$\,$$n_{H}$$\,$$\leq$$\,$2, respectively. In recent years, the explanation of the AHE based on the Berry phase theory, was used to describe the AHE in Ga$_{1\textrm{-}x}$Mn$_{x}$As crystals with a metallic type of conductivity \cite{Jungwirth2002a}. The topological explanation of the AHE was also employed theoretically for IV-VI semiconductors.\cite{Dyrdal2008a, Dyrdal2009a} The Berry phase theory predicts the square resistivity tensor component dependence $\rho_{xy}$$\,$$\propto$$\,$$\rho_{xx}^{2}$. In view of the fact that the AHE theories predict a quadratic scaling relation for two physical mechanisms leading to the formation of AHE, their differentiation (by making the scaling analysis of the experimental data) is not possible.  \\ \indent Further analysis of the observed AHE was based on the scaling analysis of the resistivity components, ie. scaling relationship given by the following equation
\begin{equation}\label{EqAHEScal01}
    \rho_{xy}(B) = R_{H} B + c_{H} \rho_{xx}^{n_{H}} M,
\end{equation}
where $c_{H}$ and $n_{H}$ are the scaling coefficients. This analysis enabled the assessment of the dominant scattering mechanisms responsible for the observed AHE. Scaling relation of the AHE was solved by fitting the experimental results to the Eq.$\;$\ref{EqAHEScal01}. The data analysis was performed with the same assumption about the normal Hall coefficient $R_{H}$ as in the previous series of fits. The $c_{H}$ and $n_{H}$ constants were taken as fitting parameters. The selected results of the fitting procedure, together with the experimental data, are presented in Fig.$\;$\ref{FigAHEScal}.
Analysis of the results indicates a good agreement (with variance smaller than 10$^{-12}$) between the experimental data and the theoretical curves given by Eq.$\;$\ref{EqAHEScal01}. As a result of the fitting  procedure we have estimated the temperature dependence of the $n_{H}$ scaling coefficient. Due to the high complexity of this analysis, it was not possible to obtain a smooth temperature dependence of $n_{H}$. The obtained values of $n_{H}$ are contained in the region between 1.1 and 1.3 for all our samples (exemplary values are presented in Table$\;$\ref{TabAHEFitRes} and Fig.$\;$\ref{FigAHEScal}). The theories of topological AHE predict $n_{H}$$\,$$\,$$=$$\,$2 for Berry phase intrinsic mechanism, and our values of $n_{H}$ are far from 2. The values of $n_{H}$ indicate that the AHE in our Ge$_{1\textrm{-}x\textrm{-}y}$Sn$_{x}$Mn$_{y}$Te samples was dominated by the extrinsic skew scattering processes. However, the presence of other scattering mechanisms giving a small contribution to AHE is also evident in our samples. It should be noted, that in the case of crystals with a high Mn content in the alloy the smaller values of $n_{H}$ were obtained. It might signify that in the high Mn-content samples the skew scattering mechanism becomes even more pronounced.

\section{Summary}

To conclude, we have shown the results of magnetotransport studies of spin-glass-like Ge$_{1\textrm{-}x\textrm{-}y}$Sn$_{x}$Mn$_{y}$Te samples with chemical composition 0.083$\,$$\leq$$\,$$x$$\,$$\leq$$\,$0.142 and 0.012$\,$$\leq$$\,$$y$$\,$$\leq$$\,$0.119. Our previous investigations showed that the spin-glass-like state appears at temperatures lower than 60$\;$K. \\ \indent The high-field magnetotransport studies show the presence of negative magnetoresistance in the studied alloy at $T$$\,$$<$$\,$$T_{SG}$, with magnitude of the magnetoresistance being an increasing function of the Mn-content, $y$. Two mechanisms are responsible for the observed negative magnetoresistance in our samples, namely weak localization and spin-disorder scattering mechanism. A strong anomalous Hall effect displaying hysteresis was observed in all our samples at $T$$\,$$<$$\,$$T_{SG}$.  \\ \indent The AHE coefficient $R_{S}$ was found to be composition dependent, changing in the range of 10$^{-7}$$\,$$<$$\,$$R_{S}$$\,$$<$$\,$10$^{-6}$$\;$m$^{3}$/C. The scaling analysis of the AHE shows that the extrinsic skew scattering mechanism, accompanied with skew scattering, is the main physical mechanism responsible for the AHE in Ge$_{1\textrm{-}x\textrm{-}y}$Sn$_{x}$Mn$_{y}$Te crystals.

\section{Acknowledgements}

\noindent The research was supported by the Foundation for Polish Science - HOMING PLUS Programme co-financed by the European Union within European Regional Development Fund.

\end{document}